\documentclass[preprintnumbers,amsmath,amssymb,10pt,showpacs, twocolumn]{revtex4}

\usepackage{graphicx}
\usepackage{dcolumn}
\usepackage{bm}

\begin{document}

\title{Probabilistic teleportation via  a non-maximally entangled GHZ state}

\author{Yan Fengli, Yan Tao}

\affiliation { College of Physics  Science and Information
Engineering, Hebei Normal University, Shijiazhuang 050016, China
\\
Hebei Advanced Thin Films Laboratory,  Shijiazhuang 050016, China}

\date{\today}

\date{\today}

\begin{abstract}
We present  a scheme for probabilistic teleportation via a
non-maximally entangled GHZ state. Quantum teleportation will
succeed with a certain probability if the sender makes a generalized
Bell state measurement, the cooperator performs a generalized $X$
basis measurement, and the receiver introduces an auxiliary
particle, performs a collective unitary transformation and makes a
measurement on the auxiliary particle.  The success probability of
the teleportation is given.  We also obtain the maximum of the
success probability of the teleportation.
\end{abstract}


 \maketitle

 Quantum teleportation, which  allows transportation of an unknown state
 from a sender Alice  to  a spatially distant receiver Bob with the aid of the   previously shared entanglement
 and classical communication,  is  regarded as one of the most striking
 results of quantum information theory \cite {NielsenChuang}.
  It  plays an important role in the development of quantum computation and quantum communication \cite{s1,s2,s3,s4,s5,s6,BDEFMS}.

 The original protocol of Bennett et al. \cite {s1} involves teleportation of
 an arbitrary  state of a  qubit via  an Einstein-Podolsky-Rosen (EPR) pair and by transmitting
 two bits of classical information from  Alice to  Bob. Here Alice
  knows neither the state to
 be teleported nor the location of the
  intended receiver, Bob. They also presented a protocol for teleporting
  an unknown state of a qudit via a maximally entangled state in $d\times d$
  dimensional Hilbert space and by sending $2{\rm log}_2d$ bits of
  classical information.

 Since then, quantum teleportation has
  got great development \cite {Braunstein, VaidmanPRA1994, BKprl1998,   GRpra,  KB, s7, APpla, PAjob, DLLZWpra2005,Zhangzhanjun} and has
been experimentally demonstrated by several groups \cite
{BPMEWZ,FSBFKP,NKL,Boschi}. It was generalized to a more general
situation,
  where two parties may not start with a set of pure entangled states,
  but with a noisy quantum channel.
   In order to arrive at their goal of transmitting
 unknown  quantum state over this noisy quantum channel, they could
  use an error correcting code \cite {Gottesman}, or alternatively they
  can  share the entanglement through  this noisy channel first and then  use teleportation \cite {BBP96}.

Quantum teleportation is also possible for infinite dimensional
Hilbert space, for example  in position-momentum space with
continuous variable states, which is called  continuous variable
quantum teleportation \cite {Braunstein,
  VaidmanPRA1994, BKprl1998}.

Karlsson and Bourennane put forward the controlled quantum
 teleportation protocol \cite{KB,DLLZWpra2005,Zhangzhanjun,s7}.
 In the protocol, one can  perfectly transport an unknown state
  from one place to another place via previously shared
Greenberger-Horne-Zeilinger (GHZ) state by means of local operations
and classical communications  under the control  of the third party.
The signal state can not be transmitted unless the third party gives
a permission. The controlled quantum teleportation is useful in
networked quantum information processing,
 and has other interesting applications, such as
in opening account  authorized by the managers in a network.

If the quantum channel is not in  a maximally entangled state then
one cannot transport a qubit with unit fidelity and unit
probability. However, it was shown that using a non-maximally
entangled state one can have unit fidelity teleportation but with a
probability less than unit-called probabilistic quantum
teleportation \cite {APpla,PAjob}. Using a non-maximally entangled
basis as a measurement basis this was shown to be possible.
Subsequently, this probabilistic scheme has been generalized to
teleport $N$ qubits \cite {GRpra} and controlled teleportation \cite
{s7}.

 In this  paper, we will investigate
 the probabilistic teleportation via a quantum channel of a non-maximally entangled GHZ state.

  First we consider three-partite probabilistic teleportation
  protocol.

  Suppose that  Alice has a qubit in state
  \begin{equation}
|\phi\rangle_{A_1}=\alpha|0\rangle+\beta|1\rangle,  ~~~
|\alpha|^2+|\beta|^2=1
  \end{equation}
and she wants to transport this state to the receiver Bob. Of
course, the identity of $|\phi\rangle_{A_1}$ is  unknown to Alice.

Suppose that a quantum channel shared by  Alice,  Bob
  and the collaborator Charlie is a non-maximally entangled GHZ state
  \begin{equation}
|{\rm GHZ}\rangle_{A_2CB}=N(|000\rangle+n|111\rangle),  N=\frac
{1}{\sqrt {1+|n|^2}}.
  \end{equation}
Here we assume that  Charlie is cooperative and loyal.
  The particles $A_1$ and $A_2$ are in  Alice's possession,
  particle $B$ is in Bob's possession and particle $C$ belongs to
  Charlie.

  The overall state of the whole system reads
  \begin{eqnarray}
  &|\psi\rangle_{A_1A_2CB}&=|\phi\rangle_{A_1}\otimes|{\rm
  GHZ}\rangle_{A_2CB}\\\nonumber
  &&=(\alpha|0\rangle+\beta|1\rangle)N(|000\rangle+n|111\rangle).
  \end{eqnarray}

 We define the generalized Bell states \cite{GRpra}
  \begin{eqnarray}
&&\nonumber|\phi^+_m\rangle=M(|00\rangle+m|11\rangle),\\
&&\nonumber|\phi^-_m\rangle=M(m^*|00\rangle-|11\rangle),\\
&&\nonumber|\psi^+_m\rangle=M(|01\rangle+m|10\rangle),\\
&&|\psi^-_m\rangle=M(m^*|01\rangle-|10\rangle)
\end{eqnarray}
and the generalized $X$ basis
\begin{eqnarray}
&&\nonumber|+\rangle=a(|0\rangle+b|1\rangle),\\
&&|-\rangle=a(b^*|0\rangle-|1\rangle).
\end{eqnarray}
Here
\begin{equation}\nonumber
M=\frac {1}{\sqrt {1+|m|^2}}
\end{equation}
and
\begin{equation}\nonumber
a=\frac {1}{\sqrt {1+|b|^2}}.
\end{equation}
Evidently, in general the generalized Bell states are not maximally
entangled.

A simple algebraic rearrangement of Eq.(3) in terms of   the
generalized Bell states Eq.(4) and the generalized $X$ basis Eq.(5)
yields
\begin{eqnarray}
&&\nonumber~~~|\psi\rangle_{A_1A_2CB}\\
&&\nonumber=|\phi\rangle_{A_1}\otimes|{\rm GHZ}\rangle_{A_2CB}\\
&&\nonumber=NMa\{|\phi^+_m\rangle_{A_1A_2}[|+\rangle_C(\alpha|0\rangle+m^*n\beta
b^*|1\rangle)_B\\
&&\nonumber~~~~~~~~~~~~~~~~~~~~~~~+|-\rangle_C(\alpha
b|0\rangle-m^*n\beta
|1\rangle)_B]\\
&&\nonumber~~~~~~~~~~~+|\phi^-_m\rangle_{A_1A_2}[|+\rangle_C(m\alpha|0\rangle-n\beta
b^*|1\rangle)_B\\
&&\nonumber~~~~~~~~~~~~~~~~~~~~~~~+|-\rangle_C(m\alpha
b|0\rangle+n\beta |1\rangle)_B]\\
&&\nonumber~~~~~~~~~~~+|\psi^+_m\rangle_{A_1A_2}[|+\rangle_C(n\alpha
b^*|1\rangle+m^*\beta|0\rangle)_B \\
&&\nonumber~~~~~~~~~~~~~~~~~~~~~~~+|-\rangle_C(m^*\beta
b|0\rangle-n\alpha |1\rangle)_B]\\
&&\nonumber~~~~~~~~~~~+|\psi^-_m\rangle_{A_1A_2}[|+\rangle_C(mn\alpha
b^*|1\rangle-\beta
|0\rangle)_B\\
&&\nonumber~~~~~~~~~~~~~~~~~~~~~~~-|-\rangle_C(mn\alpha
|1\rangle+\beta b |0\rangle)_B]\}\\\nonumber
&&\nonumber=NM\{{\sqrt
{|\alpha|^2+|mn\beta|^2}}|\phi^+_m\rangle_{A_1A_2}\\
&&\nonumber[\frac{\sqrt{|\alpha|^2+|mnb\beta|^2}}{\sqrt
{(|\alpha|^2+|mn\beta|^2)(1+|b|^2)}}|+\rangle_C\frac
{(\alpha|0\rangle+m^*n\beta
b^*|1\rangle)_B}{\sqrt{|\alpha|^2+|mnb\beta|^2}}\\
&&\nonumber+\frac {{\sqrt{|\alpha b|^2+|mn\beta|^2}}}{\sqrt
{(|\alpha|^2+|mn\beta|^2)(1+|b|^2)}}|-\rangle_C\frac {(\alpha
b|0\rangle-m^*n\beta
|1\rangle)_B}{\sqrt{|\alpha b|^2+|mn\beta|^2}}]\\
&&\nonumber+{\sqrt
{|m\alpha|^2+|n\beta|^2}}|\phi^-_m\rangle_{A_1A_2}\\
&&\nonumber[\frac {{\sqrt {|m\alpha|^2+|n\beta b|^2}}}{\sqrt
{(|m\alpha|^2+|n\beta|^2)(1+|b|^2)}}|+\rangle_C\frac
{(m\alpha|0\rangle-n\beta
b^*|1\rangle)_B}{\sqrt {|m\alpha|^2+|n\beta b|^2}}\\
&&\nonumber+\frac {{\sqrt {|m\alpha b|^2+|n\beta|^2}}}{\sqrt
{(|m\alpha|^2+|n\beta|^2)(1+|b|^2)}}|-\rangle_C\frac{(m\alpha
b|0\rangle+n\beta
|1\rangle)_B}{\sqrt {|m\alpha b|^2+|n\beta|^2}}]\\
&&\nonumber+{\sqrt
{|n\alpha|^2+|m\beta|^2}}|\psi^+_m\rangle_{A_1A_2}
\\&&\nonumber[\frac{{\sqrt {|n\alpha b|^2+|m\beta|^2}}}{\sqrt
{(|n\alpha|^2+|m\beta|^2)(1+|b|^2)}}|+\rangle_C\frac {(n\alpha
b^*|1\rangle+m^*\beta|0\rangle)_B}{\sqrt {|n\alpha b|^2+|m\beta|^2}} \\
&&\nonumber+\frac{{\sqrt{|m\beta b|^2+|n\alpha|^2}}}{\sqrt
{(|n\alpha|^2+|m\beta|^2)(1+|b|^2)}}|-\rangle_C\frac {(m^*\beta
b|0\rangle-n\alpha |1\rangle)_B}{\sqrt{|m\beta b|^2+|n\alpha|^2}}]\\
&&\nonumber+{\sqrt
{|mn\alpha|^2+|\beta|^2}}|\psi^-_m\rangle_{A_1A_2}
\\
&&\nonumber[\frac {{\sqrt {|mn\alpha b|^2+|\beta|^2}}}{\sqrt
{(|mn\alpha|^2+|\beta|^2)(1+|b|^2)}}|+\rangle_C\frac {(mn\alpha
b^*|1\rangle-\beta
|0\rangle)_B}{\sqrt {|mn\alpha b|^2+|\beta|^2}}\\
&&\nonumber-\frac{\sqrt{|mn\alpha|^2+|\beta b|^2}}{\sqrt
{(|mn\alpha|^2+|\beta|^2)(1+|b|^2)}}|-\rangle_C
\frac {(mn\alpha |1\rangle+\beta b |0\rangle)_B}{\sqrt{|mn\alpha|^2+|\beta b|^2}}]\}.\\
\end{eqnarray}

In virtue of Eq.(6), the probabilistic teleportation can be
accomplished by the following steps. Firstly, Alice performs a
generalized Bell state measurement on qubits $A_1, A_2$ shown in
Eq.(4). Evidently, $|\phi^+_m\rangle,$ $|\phi^-_m\rangle$,
$|\psi^+_m\rangle$, and $|\psi^-_m\rangle$ will occur with
probabilities $N^2M^2(|\alpha|^2+|mn\beta|^2)$,
 $N^2M^2(|m\alpha|^2+|n\beta|^2)$,
$N^2M^2(|n\alpha|^2+|m\beta|^2)$,  and
$N^2M^2(|mn\alpha|^2+|\beta|^2)$,
 respectively. Then Alice communicates to the
collaborator Charlie and Bob the outcome of the generalized Bell
state measurement and the value of $m$. Later on  Charlie performs a
generalized $X$
 basis measurement stated in Eq.(5) on his qubit $C$.
After that Charlie  tells Bob his measurement outcome and the value
of $b$. The resulting states of Bob's qubit will be respectively
\begin{equation}
\frac {(\alpha|0\rangle+m^*n\beta
b^*|1\rangle)_B}{\sqrt{|\alpha|^2+|mnb\beta|^2}}=I\frac
{(\alpha|0\rangle+m^*n\beta
b^*|1\rangle)_B}{\sqrt{|\alpha|^2+|mnb\beta|^2}},
\end{equation}
\begin{equation}
\frac {(\alpha b|0\rangle-m^*n\beta |1\rangle)_B}{\sqrt{|\alpha
b|^2+|mn\beta|^2}}=\sigma_z\frac {(\alpha b|0\rangle+m^*n\beta
|1\rangle)_B}{\sqrt{|\alpha b|^2+|mn\beta|^2}},
\end{equation}
\begin{equation}
 \frac
{(m\alpha|0\rangle-n\beta b^*|1\rangle)_B}{\sqrt
{|m\alpha|^2+|n\beta b|^2}}=\sigma_z\frac {(m\alpha|0\rangle+n\beta
b^*|1\rangle)_B}{\sqrt {|m\alpha|^2+|n\beta b|^2}},
\end{equation}
\begin{equation}
\frac{(m\alpha b|0\rangle+n\beta |1\rangle)_B}{\sqrt {|m\alpha
b|^2+|n\beta|^2}}=I\frac{(m\alpha b|0\rangle+n\beta
|1\rangle)_B}{\sqrt {|m\alpha b|^2+|n\beta|^2}},
\end{equation}
\begin{equation}\frac {(n\alpha
b^*|1\rangle+m^*\beta|0\rangle)_B}{\sqrt {|n\alpha
b|^2+|m\beta|^2}}=\sigma_x\frac {(n\alpha
b^*|0\rangle+m^*\beta|1\rangle)_B}{\sqrt {|n\alpha
b|^2+|m\beta|^2}},
\\\end{equation}
\begin{equation}\frac {(m^*\beta b|0\rangle-n\alpha
|1\rangle)_B}{\sqrt{|m\beta b|^2+|n\alpha|^2}}=i\sigma_y\frac
{(n\alpha |0\rangle+m^*\beta b|1\rangle)_B}{\sqrt{|m\beta
b|^2+|n\alpha|^2}},\end{equation}
\begin{equation}\frac {(mn\alpha b^*|1\rangle-\beta |0\rangle)_B}{\sqrt
{|mn\alpha b|^2+|\beta|^2}}=-i\sigma_y\frac {(mn\alpha
b^*|0\rangle+\beta |1\rangle)_B}{\sqrt {|mn\alpha
b|^2+|\beta|^2}},\end{equation}
\begin{equation}\frac {(mn\alpha |1\rangle+\beta b |0\rangle)_B}
{\sqrt{|mn\alpha|^2+|\beta b|^2}}= \sigma_x\frac {(mn\alpha
|0\rangle+\beta b |1\rangle)_B}{\sqrt{|mn\alpha|^2+|\beta b|^2}}.
\end{equation}
Here $I$ is the  two-dimensional identity, and  $\sigma_x, \sigma_y,
\sigma_z$ are the Pauli matrices. Depending on Alice's and Charlie's
measurement results Bob will be able to apply the corresponding
unitary operator to his particle $B$, transforming it to one of the
states $ \frac {(\alpha|0\rangle+m^*n\beta
b^*|1\rangle)_B}{\sqrt{|\alpha|^2+|mnb\beta|^2}}$, $\frac {(\alpha
b|0\rangle+m^*n\beta |1\rangle)_B}{\sqrt{|\alpha
b|^2+|mn\beta|^2}}$, $ \frac {(m\alpha|0\rangle+n\beta
b^*|1\rangle)_B}{\sqrt {|m\alpha|^2+|n\beta b|^2}}$,
 $\frac{(m\alpha
b|0\rangle+n\beta |1\rangle)_B}{\sqrt {|m\alpha b|^2+|n\beta|^2}}$,
$\frac {(n\alpha b^*|0\rangle+m^*\beta|1\rangle)_B}{\sqrt {|n\alpha
b|^2+|m\beta|^2}}$,
 $\frac {(n\alpha |0\rangle+m^*\beta
b|1\rangle)_B}{\sqrt{|m\beta b|^2+|n\alpha|^2}},\frac {(mn\alpha
b^*|0\rangle+\beta |1\rangle)_B}{\sqrt {|mn\alpha b|^2+|\beta|^2}}$,
 $\frac {(mn\alpha |0\rangle+\beta b
|1\rangle)_B}{\sqrt{|mn\alpha|^2+|\beta b|^2}}.$

For the state  $\frac {(\alpha c|0\rangle +\beta
d|1\rangle)_B}{\sqrt {|\alpha c|^2+|\beta d|^2}}$, in which  $c$ and
$d$ are known, but $\alpha$ and $\beta$ are unknown to us, it is
possible to obtain the state ${\alpha |0\rangle +\beta |1\rangle}$
with a certain probability by the following method.

    In order to achieve the above purpose, Bob needs to perform a unitary transformation
    \begin{equation}
    U=\left (\begin{array}{cc}\exp(-i\arg c)& 0\\
    0&\exp(-i\arg d)\end{array}\right)\end{equation}
     on particle $B$ under the basis
    $\{|0\rangle, |1\rangle\}$, then  particle $B$ is in the state
    \begin{equation}\frac {(\alpha |c||0\rangle +\beta |d||1\rangle)_B}{\sqrt
{|\alpha c|^2+|\beta d|^2}}.\end{equation}
 Without loss of
generality we assume that $|c| < |d|$. Furthermore, Bob needs to
introduce an
    auxiliary particle $D$ with the initial state $|0\rangle_D$ and
    performs a collective unitary transformation
  \begin{equation}U_{BD}=\left(%
\begin{array}{cccc}
  1 & 0 & 0 & 0 \\
  0 & 1 & 0 & 0 \\
  0 & 0 & |\frac {c}{d}| & \sqrt{1-|\frac {c}{d}|^2} \\
  0 & 0 & -\sqrt{1-|\frac {c}{d}|^2} & |\frac {c}{d}| \\
\end{array}%
\right)\end{equation} on  particles $B$ and $D$ under the basis
$\{|00\rangle_{BD}, |01\rangle_{BD}, |10\rangle_{BD},
|11\rangle_{BD}\}$. After that the state of the particles $B$ and
$D$ becomes
\begin{eqnarray}
&&U_{BD}\frac {(\alpha |c||0\rangle +\beta |d||1\rangle)_B}{\sqrt
{|\alpha c|^2+|\beta d|^2}}|0\rangle_D\\\nonumber
 &=&\frac {|c|(\alpha
|0\rangle +\beta |1\rangle)_B|0\rangle_D -\beta\sqrt
{|d|^2-|c|^2}|1\rangle_B|1\rangle_{D}}{\sqrt {|\alpha c|^2+|\beta
d|^2}}.\end{eqnarray} Then a measurement on Bob's auxiliary particle
$D$ in the basis $\{|0\rangle_D,|1\rangle_D\}$  follows. If
$|0\rangle_D$ occurs,  we obtain the state ${\alpha |0\rangle +\beta
|1\rangle}$ with probability $ \frac {|c|^2}{|\alpha c|^2+|\beta
d|^2}$, i.e. the teleportation is successful. Otherwise the
teleportation fails.

Based on the above argument, it is easy to image that the final step
of the teleportation protocol is to introduce an auxiliary particle
and make a collective unitary on the signal particle and the
auxiliary particle. Then Bob performs a measurement on the auxiliary
particle.

By Eq.(6), it is easy to show that the success probability of the
teleportation is
\begin{eqnarray}
 \nonumber
 &P=&2N^2M^2a^2({\min\{1,|mnb|^2\}}+{\min\{|b|^2,
|mn|^2\}}\\\nonumber & &+
{\min\{|m|^2,|nb|^2\}}+{\min\{|n|^2,|mb|^2\}}).\\
\end{eqnarray}

Let us define
\begin{equation}\begin{array}{ccc}
\xi=N^2,& \zeta=M^2,& \eta=a^2.\end{array}
\end{equation}
Then we obtain
\begin{eqnarray}
&P=&2(\min\{\xi\zeta\eta,(1-\xi)(1-\zeta)(1-\eta)\}\\\nonumber
&&+\min\{\xi\zeta(1-\eta), (1-\xi)(1-\zeta)\eta\}\\\nonumber &&+
\min\{\xi(1-\zeta)\eta,(1-\xi)\zeta(1-\eta)\}\\\nonumber
&&+\min\{(1-\xi)\zeta\eta,\xi(1-\zeta)(1-\eta)\}).
\end{eqnarray}

Obviously,
\begin{equation}
P(\xi,\zeta,\eta)=P(1-\xi,\zeta,\eta)=P(\xi,1-\zeta,\eta)=P(\xi,\zeta,1-\eta).
\end{equation}
Without loss of generality, we suppose
\begin{equation}
\begin{array}{ccc}
0<\xi\leq \frac {1}{2}, & 0<\zeta\leq \frac {1}{2},& 0<\eta\leq
\frac {1}{2}. \end{array}
\end{equation}
Therefore
\begin{eqnarray}
&P=&2(\xi\zeta\eta +\min\{\xi\zeta(1-\eta),
(1-\xi)(1-\zeta)\eta\}\\\nonumber &&+
\min\{\xi(1-\zeta)\eta,(1-\xi)\zeta(1-\eta)\}\\\nonumber
&&+\min\{(1-\xi)\zeta\eta,\xi(1-\zeta)(1-\eta)\}).
\end{eqnarray}

 Next we will find  the  maximum of $P$ when $\xi$ is fixed.

Since
\begin{equation}
P(\xi,\zeta,\eta)=P(\xi,\eta,\zeta),
\end{equation}
so the maximum of success probability $P$ must occur in the region
\begin{equation}\label{ll}
0<\zeta\leq \eta\leq\frac {1}{2}.
\end{equation}

Assume
\begin{equation}
 \{\begin{array}{l}\xi\zeta(1-\eta)> (1-\xi)(1-\zeta)\eta,\\
\xi(1-\zeta)\eta> (1-\xi)\zeta(1-\eta), \end{array}\end{equation} we
have
\begin{equation}
\xi^2>(1-\xi)^2,
\end{equation}
which does not  hold as $\xi\leq \frac {1}{2}$. So the hypothesis
Eq.(27) is wrong. Similarly we can prove that
\begin{equation}
\{\begin{array}{l}
\xi(1-\zeta)\eta> (1-\xi)\zeta(1-\eta),\\
(1-\xi)\zeta\eta> \xi(1-\zeta)(1-\eta),
\end{array}\end{equation}
is not holed also.

 Evidently
\begin{equation}\label{*}
\{\begin{array}{l}\xi\zeta(1-\eta)> (1-\xi)(1-\zeta)\eta,\\
\xi(1-\zeta)\eta< (1-\xi)\zeta(1-\eta), \end{array}\end{equation} is
wrong, because if Eq.(\ref{*}) holds, we must have
\begin{equation}
 (1-\eta)/\eta>(1-\zeta)/\zeta
\end{equation}
 that  means $\zeta> \eta$, which contradicts with Eq.(\ref{ll}).

 Therefore, the whole region indicated by Eq.(\ref{ll}) can be divided into the following three regions $E$, $F$,
 $G$. We will discuss the maximal success probability in each
 region.

 1. The region $E$  is defined by
\begin{equation}
\begin{array}{l}\xi\zeta(1-\eta)\leq (1-\xi)(1-\zeta)\eta,\\
\xi(1-\zeta)\eta\leq (1-\xi)\zeta(1-\eta),\\
(1-\xi)\zeta\eta\leq \xi(1-\zeta)(1-\eta).
\end{array}\end{equation}

In this region, the success probability
\begin{equation} P=2[\xi(\eta+\zeta)+(1-2\xi)\zeta\eta].
\end{equation}
 One can easily deduce  that the maximum of $P$ in this region must
occur in the boundary  of  $E$.

2. The region $F$ satisfies
\begin{equation}
\begin{array}{l}\xi\zeta(1-\eta)\leq (1-\xi)(1-\zeta)\eta,\\
\xi(1-\zeta)\eta\leq (1-\xi)\zeta(1-\eta),\\
(1-\xi)\zeta\eta\geq \xi(1-\zeta)(1-\eta).
\end{array}\end{equation}
In this region, we have
\begin{equation}
\xi\leq \zeta, \eta.
\end{equation}

It is straightforward  to obtain the success probability
\begin{equation} P=2\xi.\end{equation}
in region $F$.

  3. The region $G$ satisfies
\begin{equation}
\begin{array}{l}\xi\zeta(1-\eta)\leq (1-\xi)(1-\zeta)\eta,\\
\xi(1-\zeta)\eta\geq (1-\xi)\zeta(1-\eta),\\
(1-\xi)\zeta\eta\leq \xi(1-\zeta)(1-\eta).
\end{array}\end{equation}
In this region, we have
\begin{equation}
\zeta\leq \xi, \eta.
\end{equation}

 Obviously,
\begin{eqnarray}
&P=&2\zeta< 2\xi
\end{eqnarray}
in region $G$.

 Synthesizing all cases above, we arrive at the conclusion that  the maximum
of the success probability of the teleportation is
\begin{equation}
P_{max}=2\xi
\end{equation}
and it occurs in the region $F$.

Now we generalize the above teleportation protocol to the case of
$L$-parties. Suppose that the state Alice wants to transmit is
stated in Eq.(1), and the quantum channel shared between Alice and
other $L-1$ parties
  is
  \begin{equation}
|{\rm GHZ}\rangle_{12\cdots
L}=N(|00\cdots0\rangle+n|11\cdots1\rangle)_{12\cdots L},
  \end{equation}
where $N$ is defined by  Eq.(2). We assume that all members of
$L$-parties are cooperative and loyal, particles $A$ and $1$ belong
to Alice, particle $i$ is in  party $i$'s possession, $i=2,\cdots,
L$. The total state of the whole system can be written as
\begin{equation}
|\psi\rangle_{A12\cdots N}=|\phi\rangle_A\otimes|{\rm
GHZ}\rangle_{12\cdots N}.
\end{equation}

Without loss of generality,  we assume that Alice wants the $L$-th
party to receive the state. Firstly Alice  performs a generalized
Bell basis measurement on qubits $A$ and 1 shown  in Eq.(4) and
publishes the outcome of the measurement. After that   party $i$
makes the generalized $X$ basis measurement on his qubit $i$ and
announces the result of the measurement, $i=2,\cdots, L-1$. Finally,
according to the results of the measurements, party $L$ introduces
an auxiliary particle, implements a collective unitary
transformation on the particle $L$ and the auxiliary particle, and
performs  a measurement on the auxiliary particle. By completing the
above steps, an unknown quantum state has been transmitted with unit
fidelity to the receiver with a certain probability.

In summary, we have presented  a scheme for probabilistic
teleportation via a non-maximally entangled GHZ state. Quantum
teleportation will succeed with a certain probability if the sender
makes a generalized Bell state measurement, the cooperator performs
a generalized $X$ basis measurement, and the receiver introduces an
auxiliary particle, performs a collective unitary transformation and
makes a measurement on the auxiliary particle.  The success
probability of the teleportation is given.  We also obtain the
maximum of the success probability of the teleportation. The
teleportation scheme has also been generalized to the more general
case of $L$-parties.\\[0.2cm]

{\noindent\bf Acknowledgments}\\[0.2cm]

 This work was supported by the National Natural Science Foundation of China under Grant No: 10671054, Hebei Natural Science Foundation of China under Grant
No: 07M006, and  the Key Project of Science and Technology Research
of Education Ministry of China under Grant No: 207011.

\end{document}